\documentclass{article}
\usepackage{spconf,amsmath,graphicx,url,cite,amssymb,bm,here}

\usepackage{algorithm}
\usepackage{algorithmic}
\usepackage{comment}

\newcommand{\argmax}{\mathop{\rm arg~max}\limits}

\def\x{{\mathbf x}}
\def\h{{\mathbf h}}
\def\y{{\mathbf y}}
\def\Q{{\mathbf Q}}
\def\K{{\mathbf K}}
\def\V{{\mathbf V}}
\def\W{{\mathbf W}}
\def\R{{\mathbb R}}
\def\Z{{\mathbf Z}}
\def\u{{\mathbf u}}

\def\head{\mathrm{head}}
\def\model{\mathrm{model}}

\def\EndDetect{\mathrm{EndDetect}}

\title{Streaming Transformer ASR with Blockwise Synchronous Beam Search}
\name{Emiru Tsunoo$^1$, Yosuke Kashiwagi$^1$, Shinji Watanabe$^2$}
\address{
  $^1$Sony Corporation, Japan\\
  $^2$Johns Hopkins University, USA}
\begin{document}
\ninept
\maketitle
\begin{abstract}
The Transformer self-attention network has shown promising performance as an alternative to recurrent neural networks in end-to-end (E2E) automatic speech recognition (ASR) systems.
However, Transformer has a drawback in that the entire input sequence is required to compute both self-attention and source--target attention.
In this paper, we propose a novel blockwise synchronous beam search algorithm based on blockwise processing of encoder to perform streaming E2E Transformer ASR.
In the beam search, encoded feature blocks are synchronously aligned using a block boundary detection technique, where a reliability score of each predicted hypothesis is evaluated based on the end-of-sequence and repeated tokens in the hypothesis.
Evaluations of the HKUST and AISHELL-1 Mandarin, LibriSpeech English, and CSJ Japanese tasks show that the proposed streaming Transformer algorithm outperforms conventional online approaches, including monotonic chunkwise attention (MoChA), especially when using the knowledge distillation technique.  
An ablation study indicates that our streaming approach contributes to reducing the response time, and the repetition criterion contributes significantly in certain tasks.
Our streaming ASR models achieve comparable or superior performance to batch models and other streaming-based Transformer methods in all tasks considered.
\end{abstract}
\noindent\textbf{Index Terms}: speech recognition, end-to-end, Transformer, self-attention network, knowledge distillation

\section{Introduction}

\label{sec:intro}
End-to-end (E2E) automatic speech recognition (ASR) has been attracting attention as a method for directly integrating acoustic models and language models (LMs) because of its simple training and efficient decoding procedures. 
In recent years, various models have been studied, such as connectionist temporal classification (CTC) \cite{graves06, 
miao15, amodei16}, attention-based encoder--decoder models \cite{chorowski15, chan16, 
chiu18}, their hybrid models \cite{watanabe17}, and the RNN-transducer \cite{
graves13rnnt,rao17}.
Transformer \cite{vaswani17} has been successfully introduced into E2E ASR by replacing RNNs \cite{sperber18, salazar19, 
zhao19}, and it outperforms bidirectional RNN models in most tasks \cite{karita19}.
Transformer has multihead self-attention network (SAN) layers and source--target attention (STA) layers, which can leverage a combination of information from completely different positions of the input.

However, similarly to bidirectional RNN models \cite{schuster97}, Transformer has a drawback in that the entire utterance is required to compute the attentions, making its use in streaming ASR systems difficult.
In addition, the memory and computational requirements of Transformer grow quadratically with input sequence length, which makes it difficult to apply to long speech utterances.
These problems generally appear when we use SAN, and various works have been recently carried out to tackle these problems for SAN-based acoustic modeling, CTC, and transformer  toward streaming ASR \cite{sperber18,moritz20,povey18,dong19}.
These approaches simply introduce blockwise processing for the SAN layers.
Miao {\it et al.} \cite{miao2020} proposed using the previous chunk, inspired by Transformer XL \cite{dai19}.
Furthermore, context-aware inheritance mechanism is also proposed \cite{tsunoo19}.
In that approach, a context embedding vector handed over from the previously processed block helps encode not only local acoustic information, but also global linguistic, channel, and speaker attributes.

In addition to the aforementioned blockwise SAN processing, to realize entire streaming ASR for attention-based models, 
blockwise processing for STA networks is also required.
In \cite{moritz20}, a triggered attention mechanism was introduced to realize this.
However, it requires a complicated training procedure using CTC forced alignment.
Monotonic chunkwise attention (MoChA) \cite{chiu2017monotonic} is a popular approach to achieve online processing \cite{fan19, kim19, tsunoo2019towards,miao2020,inaguma20}.
However, MoChA degrades accuracy \cite{kim19, inaguma20}, and it is also difficult to control the latency within an acceptable range.

In this paper, we propose a novel blockwise synchronous beam search algorithm for streaming Transformer to provide an alternative to the MoChA or triggered attention based approaches.
The main idea of this algorithm is based on our newly introduced block boundary detection (BBD) technique for decoding after the contextual block encoder used in \cite{tsunoo19}.
The decoder receives encoded blocks one by one from the contextual block encoder.
Then, each block is decoded synchronously until an unreliable prediction occurs.
Predictions are evaluated on the fly using BBD, where a reliability score of each prediction is computed based on the end-of-sequence token, ``$\langle$eos$\rangle$,'' and a repetition of a token.
Once an unreliable prediction occurs, the decoder waits for the encoder to finish the next block.
The main contributions of this paper are summarized as follows.
1) A blockwise synchronous beam search algorithm using BBD is proposed, which is incorporated with the contextual block processing of the encoder in CTC/attention hybrid decoding scheme.
2) Knowledge distillation \cite{li2014, hinton2015,lu2017} is performed on the streaming Transformer, guided by the original batch Transformer.
3) The proposed streaming Transformer algorithm is compared with conventional approaches including MoChA.
The results indicate our approach outperforms them in the HKUST \cite{hkust06} and AISHELL-1 \cite{aishell17} Mandarin, LibriSpeech \cite{panayotov15} English, and CSJ \cite{csj} Japanese tasks.
4) The impact of each factor in the proposed blockwise synchronous beam search on latency is evaluated through an ablation study.

\section{Relation with Prior Work}
Among the various available approaches for streaming processing in Transformer, such as time-restricted Transformer \cite{moritz20,povey18}, Miao {\it et al.} \cite{miao2020} adopted chunkwise self-attention encoder (Chunk SAE), which was inspired by transformer XL \cite{dai19}, where not only the current chunk but also the previous chunk are used for streaming encoding.
Although this encoder is similar to that in \cite{tsunoo19, tsunoo2019towards}, in our case, not only the previous chunk but also a long history of chunks is efficiently referred to by introducing context embeddings.

Tian {\it et al.} \cite{tian20} applied a neural transducer \cite{jaitly2016} to the synchronous Transformer decoder, which decodes sequences in a similar manner to the approach proposed in this paper.
However, the synchronous Transformer has to be trained using a special forward--backward algorithm similarly to the training of a neural transducer using dynamic programming alignment.
In this paper, the proposed beam search algorithm does not require any additional training constraints.
Our general decoding algorithm is applied to the parameters as they are.
Whereas in \cite{tian20} the authors only use a $\langle$eos$\rangle$ token to synchronously shift the processing blocks, we also take into account a repetition of a token, which significantly improves performance in the LibriSpeech and CSJ tasks.

\section{Streaming Transformer ASR}
\subsection{Transformer ASR}
\label{ssec:transformer}
Our baseline Transformer ASR follows that described in \cite{karita19}, which is based on an encoder--decoder architecture. 
An encoder transforms a $T$-length speech feature sequence $\x = (x_{1},\dots,x_{T})$ to an $L$-length intermediate representation $\h = (h_{1},\dots,h_{L})$, where $L \leq T$ owing to downsampling.
Given $\h$ and previously emitted character outputs $\y_{0:i-1} = (y_{0},\dots,y_{i-1})$, a decoder estimates the next character $y_{i}$.

The encoder consists of two convolutional layers with stride $2$ for downsampling, a linear projection layer, and a positional encoding layer, followed by $N_{\mathrm{e}}$ encoder layers and layer normalization.
Each encoder layer has a multihead SAN followed by a position-wise feedforward network, both of which have residual connections.
In each SAN, attention weights are formed from queries ($\mathbf{Q} \in \R^{t_q\times d}$) and keys ($\mathbf{K} \in \R^{t_k\times d}$) and are applied to values ($\mathbf{V} \in \R^{t_v\times d}$) as
\vspace{-0.2cm}
\begin{align}
    \mathrm{Attention}(\Q,\K,\V) = \mathrm{softmax}\left(\frac{\Q\K^T}{\sqrt{d}}\right)\V, \label{eq:attention}
\end{align}
where typically $d = d_{\model}/M$ for the number of heads $M$.
We use multihead attention, denoted as the $\mathrm{MHD}(\cdot)$ function, as follows:
\begin{align}
    & \mathrm{MHD}(\Q,\K,\V) = \mathrm{Concat}(\head_{1},\dots,\head_{M})\W_O^n,
    \label{eq:mhead} \\
    & \head_{m} = \mathrm{Attention}(\Q\W_{Q,m}^n,\K\W_{K,m}^n,\V\W_{V,m}^n). \label{eq:head}
\end{align}

In \eqref{eq:mhead} and \eqref{eq:head}, the $n$th layer is computed with projection matrices $\W_{Q,m}^n \in \R^{d_{\model} \times d}$, $\W_{K,m}^n \in \R^{d_{\model} \times d}$, $\W_{V,m}^n \in \R^{d_{\model} \times d}$, and $\W_{O}^n \in \R^{Md \times d_{\model}}$.
For all the SANs in the encoder, $\Q$, $\K$, and $\V$ are the same matrices, which are the inputs of each SAN. 
The position-wise feedforward network is a stack of linear layers.

The decoder predicts the probability of the following character from the previous output characters $\y_{0:i-1}$ and the encoder output $\h$, i.e., $p(y_i|\y_{0:i-1},\h)$.
The character history sequence is converted to character embeddings.
Then, $N_{\mathrm{d}}$ decoder layers are applied, followed by linear projection and the Softmax function.
The decoder layer consists of an SAN and an STA, followed by a position-wise feedforward network.
The first SAN in each decoder layer applies attention weights to the input character sequence, where the input sequence of the SAN is set as $\Q$, $\K$, and $\V$.
Then, the subsequent STA attends to the entire encoder output sequence by setting $\K$ and $\V$ to be $\h$.

Transformer can leverage a combination of information from completely different positions of the input.
It requires the entire speech utterance for both the encoder and the decoder; thus, they are processed only after the end of the utterance, which causes a huge delay.
To realize a streaming ASR system, both the encoder and decoder have to be processed online synchronously.

\begin{figure}[t]
  \centering
  \includegraphics[width=0.9\columnwidth]{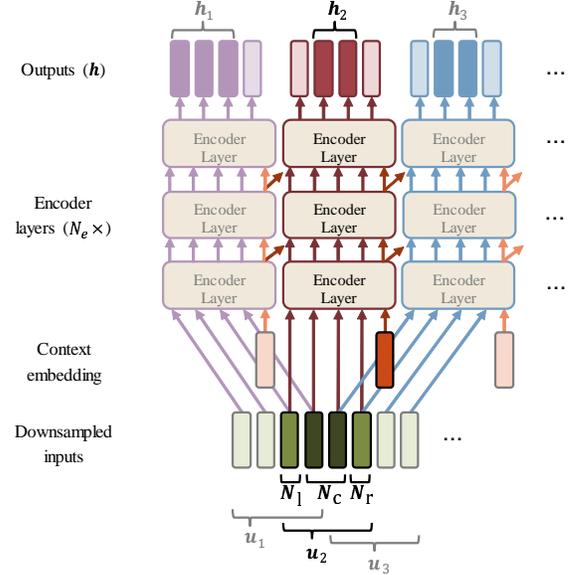}
  \vspace{-0.3cm}
  \caption{Context inheritance mechanism of the encoder}
  \label{fig:context}
  \vspace{-0.3cm}
\end{figure}

\subsection{Contextual Block Processing of the Encoder}
\label{ssec:encoder}
A simple way to process the encoder online is through blockwise computation, as in \cite{sperber18,moritz20,povey18,jaitly2016,dong19,miao2020}.
However, the global channel, speaker, and linguistic context are also important for local phoneme classification.
A context inheritance mechanism for block processing was proposed in \cite{tsunoo19} by introducing an additional context embedding vector.
As shown by the tilted arrows in Fig.~\ref{fig:context}, the context embedding vector is computed in each layer of each block and handed over to the upper layer of the following block.
Thus, the SAN in each layer is applied to the block input sequence using the context embedding vector.
A similar idea was also proposed in image and natural language processing around the same time in \cite{child19}.

Note that the blocks can overlap.
In \cite{tsunoo19}, the authors originally proposed a half-overlapping approach, where the central frames of block $b$, $\h_{b}$, are computed using the blocked input $\u_{b}$, which includes past frames as well as looking ahead for future frames.
Typically the numbers of frames used for left/center/right in \cite{tsunoo19} are $\{N_{\mathrm{l}},N_{\mathrm{c}},N_{\mathrm{r}}\}=\{4,8,4\}$, where the frames are already downsampled by a factor of $4$.
This can be easily extended to use more frames, such as $\{N_{\mathrm{l}},N_{\mathrm{c}},N_{\mathrm{r}}\}=\{16,16,8\}$, which are equivalent to the parameters in \cite{miao2020}.

\subsection{Blockwise Synchronous Beam Search of the Decoder}
The original Transformer decoder requires the entire output of the encoder $\h$.
Thus, it is not suitable for streaming processing as is.
In \cite{tsunoo2019towards}, the authors proposed using MoChA \cite{chiu2017monotonic}, which was tailored for STA.
However, accuracy significantly drops when MoChA is applied to decoder layers; this was also observed in other studies \cite{kim19, inaguma20}.
In addition, there is no guarantee that latency stays within established bounds.
To avoid these problems, we propose a novel blockwise synchronous beam search algorithm.
\subsubsection{Conventional Beam Search of Attention-based ASR}
\label{sssec:ordinary_search}
The ordinary beam search with label synchronous decoding of attention-based ASR can be formulated as a problem to find the most probable output sequence $\hat{\y}$ given all the encoded features $\h_{1:B}=(\h_{1},\dots,\h_{B})=\h$:
\vspace{-0.1cm}
\begin{align}
    \hat{\y} = \argmax_{\y\in \mathcal{V}^*}\log p(\y|\h_{1:B}),
\end{align}
where $p(\y|\h_{1:B})$ is computed by the decoder, $\mathcal{V}^*$ represents all possible output sequences, and $\hat{\y}$ is found via a beam search technique.

Let $\Omega_{i}$ be a set of partial hypotheses of length $i$, and $\Omega_{0}$ be initialized with one hypothesis with the start-of-sequence token, $y_{0}=\langle$sos$\rangle$, at the beginning of the beam search.
Until $i=I_{\rm{max}}$, each partial hypothesis in $\Omega_{i-1}$ is expanded by appending possible tokens, i.e., $\y_{0:i}=(\y_{0:i-1}, y_{i})$ where $\y_{0:i-1}$ is a partial hypothesis in $\Omega_{i-1}$.
Then, new hypotheses are stored in $\Omega_{i}$ and pruned with beam width $K$, so that only the top-$K$ scored hypotheses survive ($|\Omega_{i}|=K$).
\vspace{-0.1cm}
\begin{align}
    \Omega_{i} = \mathrm{Search}_{K}(\Omega_{i-1},\h_{1:B}) \label{eq:search}
\end{align}
The score of partial hypothesis $\y_{0:i}\in\Omega_{i}$ is accumulated in the log domain as
\vspace{-0.2cm}
\begin{align}
    \alpha(\y_{0:i}, \h_{1:B}) =\sum_{j=1}^{i}\log p(y_{j}|\y_{0:j-1}, \h_{1:B}). \label{eq:score}
\end{align}
In a conventional beam search in attention-based ASR, if $y_{i}$ is $\langle$eos$\rangle$, the hypothesis $\y_{0:i}$ is added to $\hat{\Omega}$, which denotes a set of completed hypotheses.
Finally, $\hat{\y}$ is obtained by
\vspace{-0.1cm}
\begin{align}
    \hat{\y} = \argmax_{\y\in\hat{\Omega}}\alpha(\y, \h_{1:B}).
\end{align}

\subsubsection{Blockwise Synchronous Beam Search}
\label{sssec:search}
Since the decoding problem for ASR does not depend on far-future context information, with a sufficiently high number of blocks $b (<B)$, we assume it can ignore future encoded blocks $\h_{b+1:B}$ and the following approximation is satisfied.
\vspace{-0.1cm}
\begin{align}
    \log p(y_{i}|\y_{0:i-1},\h_{1:B}) \approx \log p(y_{i}|\y_{0:i-1},\h_{1:b}) \label{eq:approx}
\end{align}
The approximation is more valid when the decoder states attend to the encoded features more locally within $b$ blocks.
Thus, the beam search is approximately carried out with the limited features encoded so far ($\h_{1:b}$).
While (\ref{eq:score}) is synchronous to output index $i$, it can be rewritten to be also synchronous to encoded block $b$, as
\vspace{-0.2cm}
\begin{align}
    \alpha(\y_{0:i}, \h_{1:B}) \approx \sum_{b=1} ^B \sum_{j=I_{b-1}+1}^{I_{b}}\log p(y_{j}|\y_{0:j-1}, \h_{1:b}), \label{eq:score2}
\end{align}
where $I_{b}$ is an index boundary, which is the last output index for the number of blocks $b$ that satisfies the approximation (\ref{eq:approx}), and $I_{0}=0$.
The main idea of this paper is to find appropriate index boundary $I_b$ during beam search.

\subsubsection{Block Boundary Detection}
\label{sssec:reliability_score}
When a hypothesis is longer than can be supported by the current encoded data, such a hypothesis is unreliable, because the approximation (\ref{eq:approx}) no longer holds.
In such cases, the decoder tends to struggle with two common errors in attention-based ASR described in \cite{watanabe17}:
\begin{enumerate}
    \item It prematurely predicts $\langle$eos$\rangle$ as the attentions reach the end of insufficient encoder blocks.
    \item It predicts a repeated token because it attends to a position that has already been attended.
\end{enumerate}
Therefore, we consider a hypothesis that contains $\langle$eos$\rangle$ or a repetition as unreliable with insufficient $b$ encoded blocks.
Further, a hypothesis that has lower score than that of the unreliable hypothesis can also be considered as unreliable. 
We propose a detection technique called BBD, where the index boundaries $I_{b}$ is found by comparing those scores on the fly.

\begin{figure}[t]
  \centering
  \includegraphics[width=1.05\columnwidth]{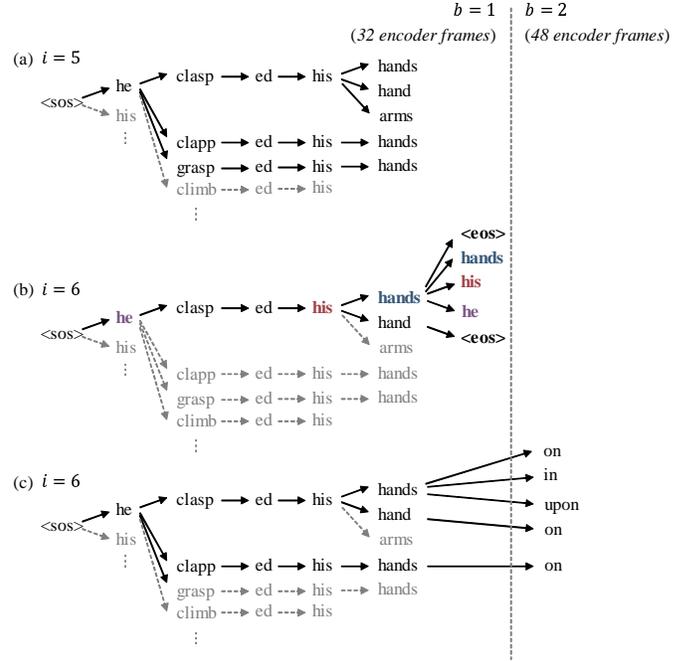}
  \vspace{-0.7cm}
  \caption{Example of the blockwise synchronous beam search of "He clasped his hands on the desk and said" with a beam width of 5}
  \label{fig:tree}
  \vspace{-0.3cm}
\end{figure}

For convenience, we share the $\langle$sos$\rangle$ token with $\langle$eos$\rangle$ ($\langle$eos$\rangle=\langle$sos$\rangle$), so that $\langle$eos$\rangle$ is also regarded as a repeat of the $\langle$sos$\rangle(=y_{0})$ token.
When token $y_{j}\in \y_{0:i-1}$ is repeatedly predicted from $\y_{0:i-1}$, the score is accumulated as $\log p(y_{j}|\y_{0:i-1}, \h_{1:b}) + \alpha(\y_{0:i-1}, \h_{1:b})$.
Thus, the highest score among unreliable hypotheses with a repetition is described as
\begin{align}
    r(\y_{0:i-1},\h_{1:b}) = \max_{0\leq j \leq i-1}\log p(y_{j}|\y_{0:i-1},\h_{1:b}) + \alpha(\y_{0:i-1}, \h_{1:b}). \label{eq:repetition}
\end{align}
As mentioned above, all the hypothesis with a lower score than $r(\y_{0:i-1},\h_{1:b})$ is considered to be unreliable.
We define a reliability score for hypothesis $\y_{0:i}$ as follows.
\vspace{-0.1cm}
\begin{align}
    s(\y_{0:i},\h_{1:b}) &= \alpha(\y_{0:i},\h_{1:b}) - r(\y_{0:i-1},\h_{1:b}) \label{eq:criterion}
\end{align}
Only when $s(\y_{0:i},\h_{1:b})>0$, the hypothesis is considered to be reliable.

As long as the encoder does not reach the end of the input utterance ($b<B$), each predicted hypothesis is evaluated using the reliability score (\ref{eq:criterion}).
If it finds an unreliable hypothesis with $s(\y_{0:i},\h_{1:b})<0$, we assume that this unreliable hypothesis $\y_{0:i}$ is already longer than current index boundary $I_b$.
Our preliminary experiments showed that, when one hypothesis $\y_{0:i}$ contains such $\langle$eos$\rangle$ or a repetition, most of the other hypotheses within the same output index $i$ also have the same tendency.
Therefore, we can empirically regard that all the hypotheses in $i$ are not considered to satisfy (\ref{eq:approx}) if at least one of the top-$K$ hypotheses is unreliable, i.e., $s(\y_{0:i},\h_{1:b})<0$.
In this way, the index boundary $I_{b}$ is assigned as the previous output index, i.e., $I_{b} = i-1$, and the decoder waits for the next block, $\h_{b+1}$, to be encoded. 
The beam search for output index $i$ resumes using hypothesis set $\Omega_{i-1}$, given encoded features $\h_{1:b+1}$.
BBD is general so that it is applicable not only to Transformer but also other architectures such as RNNs.


\subsubsection{Example of Blockwise Synchronous Beam Search}
\label{sssec:example}

Figure~\ref{fig:tree} is an example of a blockwise synchronous beam search of the decoder with beam width $K=5$.
First, the decoder starts with the first encoded block $\h_{1}$ (length is 32 when $\{N_{l}, N_{c}, N_{r}\} = \{16, 16, 8\}$).
As in Fig.~\ref{fig:tree}-(a), hypotheses are predicted from $\Omega_{4}$ with the limited encoded block and appended.
They are then stored in $\Omega_{5}$ after being pruned.


In Fig.~\ref{fig:tree}-(b), $\langle$eos$\rangle$ appears in the hypotheses, as well as repetitions (``hands,'' ``his,'' and ``he'').
In all cases, the reliability scores (\ref{eq:criterion}) are not greater than 0, because all the hypotheses in top-5 score contain repetition and the highest one is $r(\y_{0:4},\h_{1})$.
Therefore, the decoder does not store those hypotheses in $\Omega_{6}$.
Instead, the decoder waits for the encoder to output the next block $\h_{2}$ and resumes decoding from $\Omega_{5}$ using 48 encoded features $\h_{1:2}$ (Fig.~\ref{fig:tree}-(c)).
In this example, the index boundary is assigned as $I_{1}=5$.

\subsubsection{Additional Heuristics}
\label{sssec:heuristics}
Note that the same repetition will not be evaluated again with $b+1$ blocks because the repetition of a token is most likely correct when it still occurs when sufficient encoder blocks are given.
For instance, ``$\langle$sos$\rangle$ - {\it he} - clasp - ed - his - hands - {\it he}'' might be correct if it still occurs with $\h_{1:2}$.
Therefore, the hypothesis already evaluated is stored in a set, $\Omega_{\mathrm{R}}$, to prevent it from being reevaluated.
In the example in Fig.~\ref{fig:tree}, all hypotheses in (b) are stored in $\Omega_{\mathrm{R}}$.
Then, (\ref{eq:repetition}) is rewritten by excluding hypotheses in $\Omega_{\mathrm{R}}$ as
\begin{align}
\vspace{-0.2cm}
    r_{\overline{\Omega}_{\mathrm{R}}}(\y_{0:i-1},\h_{1:b}) = \max_{\substack{0\leq j \leq i-1 \\ (\y_{0:i-1}, y_{j})\in\overline{\Omega}_{\mathrm{R}}}}& \log p(y_{j}|\y_{0:i-1},\h_{1:b})\nonumber\\&+\alpha(\y_{0:i-1},\h_{1:b}).
\end{align}

The proposed beam search is carried out synchronously as the encoder finishes each block, and thus streaming decoding in Transformer is realized.
After the encoder finishes the last block $\h_{B}$, the beam search continues with all the encoded features $\h_{1:B}$ as usual until the ending criterion is met as described in \cite{watanabe17}.
The proposed beam search algorithm is summarized in Algorithm~\ref{alg:decode}.

\begin{algorithm}[t]                      
\caption{Blockwise synchronous beam search of the decoder}   
\label{alg:decode}                          
\begin{algorithmic}[1]     
\REQUIRE encoder feature blocks $\h_{b}$, total block number $B$, beam width $K$
\ENSURE $\hat{\Omega}$: complete hypotheses
\STATE \textbf{Initialize:} $y_0\gets\langle \mathrm{sos}\rangle$, $\Omega_{0} \gets \{y_{0}\}$, $\Omega_{\mathrm{R}} \gets \{\}$, $b\gets 1$, $I_*\gets I_{\mathrm{max}}$, $I_{0}\gets 0$
\WHILE{$b < B$}
\STATE {$\mathrm{NextBlock}\gets false$}
\FOR{$i\gets I_{b-1}+1$ to $I_{b}$ unless $\mathrm{NextBlock}$}
\STATE $\Omega_{i} \gets \mathrm{Search}_{K}(\Omega_{i-1},\h_{1:b})$
\FOR{{$\y_{0:i} \in \Omega_{i}$}}
\IF{$s(\y_{0:i},\h_{1:b})\leq 0$}
\STATE $\mathrm{NextBlock}\gets true$
\STATE $\Omega_{\mathrm{R}} \gets \Omega_{\mathrm{R}} \cup \y_{0:i}$ \ \ \ // store the hypothesis already evaluated
\ENDIF
\ENDFOR
\IF{$\mathrm{NextBlock}$}
\IF{$i\geq2$}
\STATE $I_{b} \gets i - 2$ \ \ \ // for conservative decoding
\ELSE
\STATE $I_{b} \gets i - 1$
\ENDIF
\STATE $b \gets b + 1$  \ \ \ // wait for the next block
\ENDIF
\ENDFOR
\ENDWHILE
\STATE // ordinary decoding follows to obtain $\hat{\Omega}$ after $b=B$
\FOR {$i\gets I_{B-1}+1$ to $I_{\mathrm{max}}$ unless $\mathrm{EndingCriterion}(\Omega_{i-1})$}
\STATE $\Omega_{i} \gets \mathrm{Search}_{K}(\Omega_{i-1},\h_{1:B})$  
\FOR{$\y_{0:i}\in\Omega_{i}$}
\IF{$y_{i}=\langle \mathrm{eos} \rangle$}
\STATE $\hat{\Omega} \gets \hat{\Omega} \cup \y_{0:i}$ 
\ENDIF
\ENDFOR
\ENDFOR
\RETURN $\hat{\Omega}$
\end{algorithmic}
\vspace{-2pt}
\end{algorithm}

More conservatively, not only $\Omega_{i}$ but also $\Omega_{i-1}$ might be considered to contain unreliable hypotheses when $s(\y_{0:i},\h_{1:b})<0$.
In this conservative case, two steps before the output index is assigned to the index boundary, i.e., $I_{b} = i-2$ instead of $I_{b} = i-1$, as in line 14 of Algorithm~\ref{alg:decode}.
In the example shown in Figure~\ref{fig:tree}, the algorithm resumes from hypothesis set $\Omega_{4}$ instead of $\Omega_{5}$.
This can reduce errors caused by the insufficient encoded features.
However, it leads to more overlaps in the decoding process, which reduces computationally efficiency.
The effectiveness of conservative decoding is evaluated in our ablation study in Sec.~\ref{ssec:ablation}.

\subsubsection{On-the-fly CTC Prefix Scoring}
Decoding is carried out jointly with CTC as in \cite{watanabe17}.
Originally, for each hypothesis, the CTC prefix score is computed as
\vspace{-1mm}
\begin{align}
    p_{\mathrm{ctc}}(\y_{0:i}|\h) = \gamma_{T}^{(\mathfrak{N})}(\y_{0:i-1}) + \gamma_{T}^{(\mathfrak{B})}(\y_{0:i-1}),
\end{align}
where the superscripts $(\mathfrak{N})$ and $(\mathfrak{B})$ denote CTC paths ending with a nonblank or blank symbol, respectively.
Thus, the entire encoded features $\h$ is required for accurate computation.
However, in the case of a blockwise synchronous beam search, computations are carried out with a limited input length.
Therefore, the CTC prefix score is computed from the blocks that are already encoded as follows:
\vspace{-2mm}
\begin{align}
\label{eq:ctc}
    p_{\mathrm{ctc}}(\y_{0:i}|\h_{1:b}) = \gamma_{T_{b}}^{(\mathfrak{N})}(\y_{0:i-1}) + \gamma_{T_{b}}^{(\mathfrak{B})}(\y_{0:i-1}),
\end{align}
where $T_{b}$ is the last frame of the currently processed block $b$.
When a new block output $\h_{b+1}$ is emitted by the encoder, the decoder resumes the CTC prefix score computation according to Algorithm 2 in \cite{watanabe17}.
Equation~(\ref{eq:ctc}) incurs a higher computational cost as the input sequence becomes long.
However, it can be efficiently computed using a technique described in \cite{seki19}.

\subsection{Knowledge Distillation Training}
\label{ssec:kd-train}
Our preliminary experiments show that parameters trained for the ordinary batch decoder perform well without significant degradation when they are directly used in the blockwise synchronous beam search of the decoder.
Therefore, instead of using special dynamic programming or a forward--backward training method as in \cite{jaitly2016, tian20}, we propose applying knowledge distillation \cite{li2014, hinton2015,lu2017} to the streaming Transformer, guided by the ordinary batch Transformer model for further improvement.

Let $q^{\mathsf{tchr}}(y_{i}|\y_{0:i-1},\h)$ be a probability distribution computed by a teacher batch model trained with the same dataset, and $p(y_{i}|\y_{0:i-1},\h)$ be a distribution predicted by the student streaming Transformer model.
The latter is forced to mimic the former distribution by minimizing the cross-entropy, which can be written as 
\vspace{-2mm}
\begin{align}
    \mathcal{L}_{\mathrm{KD}} = -\sum_{y_{i} \in \mathcal{V}} q^{\mathsf{tchr}}(y_{i}|\y_{0:i-1},\h) \log p(y_{i}|\y_{0:i-1},\h),
\end{align}
where $\mathcal{V}$ is a set of vocabulary.
The aggregated loss function for the attention encoder and decoder is calculated as 
\vspace{-1mm}
\begin{align}
    \mathcal{L}_{\mathrm{att},\mathrm{KD}} = (1-\lambda_{\mathrm{KD}}) \mathcal{L}_{\mathrm{att}} + \lambda_{\mathrm{KD}} \mathcal{L}_{\mathrm{KD}},
\end{align}
where $\lambda_{\mathrm{KD}}$ is a controllable parameter; typically $\lambda_{\mathrm{KD}}=0.5$.
Then, this loss is combined with CTC loss as in \cite{watanabe17}.

Note that the knowledge distillation is only applied to the encoder--decoder, i.e., the CTC part for the student is trained by its own.
Incorporating with training of the CTC student model would requires a complicated dynamic-programming-like matching algorithm, which is beyond the scope of this paper and left for future work.

\section{Experiments}
\subsection{Experimental Setup}
\label{ssec:setups}
We carried out experiments using the HKUST \cite{hkust06} and AISHELL-1 \cite{aishell17} Mandarin tasks, the English LibriSpeech dataset \cite{panayotov15}, and the Japanese CSJ dataset \cite{csj}.
The input acoustic features were 80-dimensional filter bank features and the pitch.

\begin{table}[t]
  \caption{CERs in the HKUST task}
  \label{tab:hkust}
  \centering
  \scalebox{0.9}{
  \begin{tabular}{l|cc}
    \hline
     & Dev  & Test \\
    \hline\hline
    \multicolumn{2}{l}{Batch processing}  \\
    \hline
    Transformer \cite{karita19} (reprod.) & 24.0 & 23.5 \\
    \ \ \ \ + {\it SpecAugment}  & 21.2 & 21.4 \\
    Chunk SAE + Batch Dec. \cite{miao2020} (reprod.) & 25.8 & 25.0 \\
    CBP-ENC + Batch Dec. \cite{tsunoo19} & 25.3 & 24.6 \\
    \ \ \ \ + {\it SpecAugment}  & 22.3 & 22.1 \\
    \hline
    \multicolumn{2}{l}{Streaming processing} \\
    \hline
    CIF + Chunk-hopping \cite{dong20} &-- & 23.6 \\
    CBP-ENC + MoChA Dec. \cite{tsunoo2019towards}&&\\
    \ \ \ \ + {\it SpecAugment} & 28.1 & 26.1 \\
    CBP-ENC + BBD (proposed) &&\\
    \ \ \ \ + {\it SpecAugment} & {22.6} & {22.6}\\
    \ \ \ \ + {\it Knowledge Distillation} & {\bf 22.2} & {\bf 22.4}\\
    \hline
  \end{tabular}
  }
  \vspace{-0.4cm}
\end{table}

\begin{table}[t]
  \caption{CERs in the AISHELL-1 task}
  \label{tab:aishell}
  \centering
  \scalebox{0.9}{
  \begin{tabular}{l|cc}
    \hline
     & Dev  & Test \\
    \hline\hline
    \multicolumn{2}{l}{Batch processing}  \\
    \hline
    Transformer ($N_{e}=6$) \cite{karita19} (reprod.) & 7.4 & 8.1 \\ 
    CBP-ENC + Batch Dec. ($N_{e}=6$) \cite{tsunoo19} & 7.6 & 8.4 \\
    CBP-ENC + Batch Dec. ($N_{e}=12$) \cite{tsunoo19} & 6.4 & 7.2 \\ 
    \hline
    \multicolumn{2}{l}{Streaming processing} \\
    \hline
    RNN-T \cite{tian19} & 10.1 & 11.8 \\
    Sync-Transformer ($N_{e}=6$) \cite{tian20} & {7.9} & 8.9 \\
    CBP-ENC + MoChA Dec. \cite{tsunoo2019towards}  & 9.7 & 9.7 \\ 
    CBP-ENC + BBD ($N_{e}=6$, proposed) & {7.6} & 8.5 \\ 
    \ \ \ \ + {\it Knowledge Distillation} & {7.6} & {8.4} \\ 
    CBP-ENC + BBD ($N_{e}=12$, proposed) & {\bf 6.4} & {\bf 7.3} \\ 
    \hline
  \end{tabular}
  }
  \vspace{-0.4cm}
\end{table}

For the training process, we used multitask learning with CTC loss as in \cite{watanabe17,karita19} with a weight of 0.3.
A linear layer was added to the encoder to project $\h$ onto the character probability for CTC. 
The Transformer models were trained 
using the Adam optimizer and Noam learning rate decay as in \cite{vaswani17}.
Decoding was performed alongside CTC, using the proposed beam search algorithm under the conservative condition described in Sec.~\ref{sssec:heuristics}.

The encoder had $N_{\mathrm{e}}=12$ layers with 2048 units and the decoder had $N_{\mathrm{d}}=6$ layers with 2048 units.
We set $d_{model}=256$ and $M=4$ for the multihead attentions. 
The input block was overlapped with parameters $\{N_{\mathrm{l}},N_{\mathrm{c}},N_{\mathrm{r}}\}=\{16,16,8\}$ to enable a comparison with \cite{miao2020}, as explained in Sec.~\ref{ssec:encoder}.
We trained the contextual block processing encoder (CBP-ENC) with the batch decoder.
The parameters for the batch decoder were directly used in the proposed blockwise synchronous beam search algorithm of the decoder using BBD for inference.

Training was carried out using ESPNet \footnote{The training and inference implementations are publicly available at \url{https://github.com/espnet/espnet}.} \cite{watanabeespnet} with the PyTorch backend.


\subsection{ASR Results}
\subsubsection{HKUST}
We used 3655 character classes with a CTC weight of 0.3 and a beam width of 10.
Shallow fusion of a two-layer LSTM LM with 650 units was applied with a weight of 0.3.
For comparison, we implemented Chunk SAE \cite{miao2020}, which is similar to our CBP-ENC approach except that it does not use the contextual embedding procedure introduced in Section \ref{ssec:encoder}.
Though we were unable to reproduce the original score in \cite{miao2020}, the implemented model performed reasonably well.

The results are listed in Table~\ref{tab:hkust}.
By comparing CBP-ENC with Chunk SAE, we can confirm that our contextual embedding approach performed better, in both cases where the batch decoder was used.
SpecAugment \cite{park19} resulted in further improvement.
For streaming processing, we obtained better performance by combining CBP-ENC and BBD rather than CBP-ENC and the MoChA decoder \cite{tsunoo2019towards}.
The knowledge distillation training in Sec.~\ref{ssec:kd-train} further improved its performance.
The proposed method achieved state-of-the-art performance as a streaming E2E approach.

\begin{table}[t]
  \caption{WERs in the LibriSpeech task (Beam width is 30)}
  \label{tab:librispeech}
  \centering
  \scalebox{0.9}{
  \begin{tabular}{l|cccc}
    \hline
     & Dev & & Test & \\
     & clean & other & clean & other \\
    \hline\hline
    \multicolumn{2}{l}{Batch processing}  \\
    \hline
    ContextNet \cite{han2020} (SOTA) & 2.1 & 4.6 & 1.9 & 4.1 \\  
    Transformer \cite{karita19} & 2.2 & 5.6 & 2.6 & 5.7 \\  
    Transformer \cite{karita19} (reprod.) & 2.5 & 6.3 & 2.8 & 6.4 \\  
    \ \ \ {\it w/ Transforemr LM} & 2.4 & 5.9 & 2.7 & 6.1 \\  

    CBP-ENC + Batch Dec. \cite{tsunoo19} & 2.7 & 7.2 & 2.9 & 7.3 \\ 
    \hline
    \multicolumn{2}{l}{Streaming processing} \\
    \hline
    CBP-ENC + CTC \cite{tsunoo19} & 3.2 & 9.0 & 3.3 & 9.1 \\ 
    CIF + Chunk-hopping \cite{dong20} & -- & -- & 3.3 & 9.6 \\
    Triggered Attention \cite{moritz20} (large, SOTA) & {2.6} & {7.2} & {2.8} & {7.3} \\ 
    CBP-ENC + BBD (proposed) & {2.5} & {6.8} & {2.7} & {7.1} \\ 
    \ \ \ {\it w/ Transformer LM} & {\bf 2.3} & {\bf 6.5} & {\bf 2.6} & {\bf 6.7} \\ 
    \hline
  \end{tabular}
  }
  \vspace{-0.4cm}
\end{table}

\begin{table}[t]
  \caption{CERs in the CSJ task}
  \label{tab:csj}
  \centering
  \scalebox{0.9}{
  \begin{tabular}{l|ccc}
    \hline
     & eval 1 & eval 2 & eval 3 \\
    \hline\hline
    \multicolumn{2}{l}{Batch processing}  \\
    \hline
    Transformer \cite{karita19} (reprod.) & 5.0 & 3.7 & 4.1 \\
    CBP-ENC + Batch Dec \cite{tsunoo19} & 5.3 & 4.0 & 4.5 \\
    \hline
    \multicolumn{2}{l}{Streaming processing} \\
    \hline
    CBP-ENC + CTC \cite{tsunoo19} & 6.2 & 4.5 & 5.2 \\
    CBP-ENC + BBD (proposed) & {\bf 5.3} & {\bf 4.1} & {\bf 4.5} \\

    \hline
  \end{tabular}
  }
  \vspace{-0.4cm}
\end{table}

\begin{table*}[t]
  \caption{Ablation study and computational speed comparison with C++ CPU implementation (Beam width is 10)}
  \label{tab:speed}
  \centering
  \scalebox{0.9}{
  \begin{tabular}{l|ccc|ccc|ccc}
 & \multicolumn{3}{c|}{HKUST}  & \multicolumn{3}{c|}{CSJ} & \multicolumn{3}{c}{Librispeech (large LM)} \\
    & CER & RTF & Response & CER (eval1/eval2/eval3) & RTF & Response & WER (clean/other) & RTF & Response  \\
  \hline
  Average utterance length & \multicolumn{3}{c|}{4.9s}  & \multicolumn{3}{c|}{4.9s} & \multicolumn{3}{c}{9.1s} \\
  \hline
  Batch Transformer \cite{karita19} (reprod.) & 21.4 & 0.07 & 0.31s & 5.0\% / 3.7\% / 4.1\%  & 0.16 & 0.74s & 2.9\% / 6.7\% & 0.36 & 3.49s  \\ 
  CBP-ENC + Batch Dec \cite{tsunoo19} & 22.1 & 0.08 & 0.31s & 5.3\% / 4.0\% / 4.5\%  & 0.17 & 0.71s & 2.8\% / 7.4\% & 0.33 & 2.81s  \\ 
  \hline
  CBP-ENC + BBD (proposed) & 22.4\% & 0.09 & 0.23s &  5.3\% / 4.1\% / 4.5\% & 0.17 & 0.52s & 3.0\% / 7.8\% & 0.35 & 1.19s  \\ 
  - conservative decoding & 22.8\% & 0.09 & 0.19s & 5.5\% / 4.2\% / 4.8\% & 0.17 & 0.50s & 5.3\% / 10.6\% & 0.35 & 1.08s  \\ 
  - repetition & 25.4\% & 0.08 & 0.15s & 28.6\% / 29.5\% / 24.8\% & 0.17 & 0.32s & 32.3\% / 39.8\% & 0.35 & 0.71s \\  
  \hline
  \end{tabular}
  }
  \vspace{-0.4cm}
\end{table*}

\subsubsection{AISHELL-1}
For this task, 4231 character classes were used with parameters \{CTC weight, beam width, LM weight\} = \{0.5, 10, 0.7\}.
To make a comparison with Sync-Transformer \cite{tian20} possible, we trained a smaller Transformer with $N_{\mathrm{e}}=6$.
The results are shown in Table~\ref{tab:aishell}.
Additionally, the results for RNN-T evaluated in \cite{tian19} are listed.
As can be seen in the results, our approach outperformed both the MoChA decoder and Sync-Transformer \cite{tian20}, especially when we applied the knowledge distillation.

\subsubsection{LibriSpeech}
\label{sssec:librispeech}
For LibriSpeech, we adopted byte-pair encoding (BPE) subword tokenization \cite{sennrich16}, which had 5000 token classes.
In addition to a large LM (four-layer LSTM with 2048 units), we evaluated the use of a Transformer LM (16-layer transformer LM with 2048 units and 8 heads); both were fused with a weight of 0.6.
CTC weight and beam width were set as 0.4 and 30.
SpecAugment \cite{park19} was also applied when it was trained.

The results are shown in Table~\ref{tab:librispeech}.
Though we did not use a large model as in \cite{karita19, moritz20}, we obtained similar results.
The proposed method achieved better performance than CTC decoding \cite{tsunoo19} and continuous integer-and-fire (CIF) online E2E ASR \cite{dong20}, which indicats that our blockwise synchronous beam search also works with BPE tokenization.
Even with the LSTM LM, we also achieved comparable performance to state-of-the-art streaming E2E ASR using triggered attention \cite{moritz20}, which was a model twice as large as ours.
Note that there is still room to improve accuracy, since our reproduction of \cite{karita19} was not as well tuned as the original paper.

\subsubsection{CSJ}
CSJ data had 3260 character classes.
The parameters were set as \{CTC weight, beam width, LM weight\} = \{0.3, 10, 0.3\}, and a two-layer LSTM LM with 650 units was fused.
SpecAugment \cite{park19} was used for data augmentation.
The results are shown in Table~\ref{tab:csj}.
The proposed method outperformed a CTC-based streaming approach \cite{tsunoo19}, and also did not degrade significantly from the batch Transformer.


\subsection{Ablation Study and Computational Speed Comparison}
\label{ssec:ablation}
We carried out an ablation study to evaluate how each factor contribute to both accuracy and computational efficiency.
HKUST, LibriSpeech, and CSJ were used.
To evaluate error rates, beam widths were fixed at 10.
To evaluate computational speed, we implemented the proposed beam search algorithm in C++, and subset of each task was used.
We used Intel Math Kernel Library 
to perform matrix operations with CPUs.
To avoid redundant computation in the decoder, we applied caching techniques similarly to \cite{he2019,saon2020}, which reduce the real-time factor (RTF) of RNN-T computations from 0.89 to 0.61 in \cite{saon2020}.
For LibriSpeech, a large LM (four-layer LSTM with 2048 units) was used as described in Sec.~\ref{sssec:librispeech}.
The latency during the utterance was not evaluated in this study because the alignment between the input and the output sequence was not provided.
The theoretical delay was 0.64 seconds because the encoder block shifted every 16 frames with 4-factor downsampling.
Instead, we measured the response time, which was the time required to finish decoding after the end of each utterance.
RTF and response time were measured with an 8 core 3.60 GHz Intel i9-9900K processor.

The results are shown in Table~\ref{tab:speed}.
The RTFs of the batch Transformer were smaller than those of the streaming Transformer for HKUST and CSJ, because the proposed streaming Transformer processed with overlaps.
As for LibriSpeech, the RTF of the batch Transformer was greater than streaming because utterance length were longer (9.1 s on average), which had a quadratic-order effect.
In addition, only LibriSpeech was used with a larger LM.
Therefore, its response time was greater than that for other tasks.
The response times of the streaming Transformer were shorter for all task owing to its efficient blockwise beam search.
Whereas the differences in performance were small for the HKUST and CSJ tasks, in which the utterances were generally short, the relative improvement observed for the  LibriSpeech task was significant due to the longer utterances.
When the decoding process was carried out without the conservative approach described in Sec.~\ref{sssec:heuristics}, the error rates slightly increased for the LibriSpeech because BBD failed to detect the block boundary, while the response times improved.
We also performed an ablation study to evaluate the repetition criterion by modifying (\ref{eq:repetition}) as \vspace{-0.1cm}
\begin{align}
    r'(\y_{0:i-1},\h_{1:b}) = \log p(\langle\mathrm{eos}\rangle|\y_{i-1},\h_{1:b}) + \alpha(\y_{0:i-1},\h_{1:b}),
\end{align}
which only evaluated $\langle\mathrm{eos}\rangle$ as in \cite{jaitly2016, tian20}.
The results indicate that the repetition of tokens is an important criterion for the blockwise synchronous beam search, because the error rates significantly increased without them, dramatically in the CSJ and LibriSpeech tasks.

\section{Conclusions}
We proposed a new blockwise synchronous beam search algorithm based on a blockwise processing of encoder to achieve streaming E2E Transformer ASR.
A block boundary detection technique was proposed, where a reliability score is computed based on  $\langle$eos$\rangle$ and repeated tokens in the hypotheses.
Using this technique, each prediction is judged as either reliable or unreliable using the current limited number of blocks from the encoder.
If a prediction is deemed unreliable, the decoder waits for the encoder to finish the next block.
Evaluations of the HKUST and AISHELL-1 Mandarin, LibriSpeech English, and CSJ Japanese tasks showed that the proposed streaming Transformer outperforms conventional online approaches including MoChA, especially when using the knowledge distillation technique.
The algorithm is general so that future work is to apply it also to the latest architecture such as Conformer \cite{gulati2020}.

\newpage
\bibliographystyle{IEEEbib}
\bibliography{mybib}

\end{document}